# Proton transfer in histidine-tryptophan heterodimers embedded in helium droplets

Bruno Bellina,[*] Daniel J. Merthe, and Vitaly V. Kresin

*Department of Physics and Astronomy, University of Southern California, Los Angeles, California 90089-0484, USA*

**Abstract**

We used cold helium droplets as nano-scale reactors to form and ionize, by electron bombardment and charge transfer, aromatic amino acid heterodimers of histidine with tryptophan, methyl-tryptophan, and indole. The molecular interaction occurring through an N-H···N hydrogen bond leads to a proton transfer from the indole group of tryptophan to the imidazole group of histidine in a radical cationic environment.

## I. INTRODUCTION

The field of protein stabilization and more specifically aromatic amino acid reactivity, a key feature of many complex biomolecular interactions, has been intensely investigated.[1-7] Interactions between amino acids, which occur through hydrogen-bonding or π-stacking interactions, primarily result in heterodimer formations. In the case of aromatic amino acids, these complexes are the place of particular aromatic-aromatic interactions. Histidine and tryptophan (whose side chains are imidazole and indole, respectively) are classic examples of such aromatic amino acids (Fig. 1). Their properties are known to be highly involved in important processes such as photo-chemistry, metal binding and proton transfer.[8-12] It has been shown that the presence of a π-radical cationic tryptophan embedded in a peptide can lead to strong hydrogen binding involving the indole NH bond.[13, 14]

Generally, the presence of a hydrogen bond in a complex is also key to permitting proton transfer.[15, 16] This particular reaction in aromatic molecules is a pervasive and highly active field of research and plays a significant role in the resultant reactions between radical amino acids.[17-21] Moreover, such interactions can lead to dramatic changes in the fragmentation patterns of polypeptides and other biomolecules.[22] Understanding these processes, for single molecules as well as complexes, is therefore crucial in proteomics. Despite the importance of these processes, the study of interactions between aromatic amino acids is still relatively underdeveloped.

Some studies have presented these non-covalent interactions as resulting from heterodimer configurations or from interactions with the protein backbone.[23-26] Recently S. Kumar et al.[27] reported a gas phase spectroscopic study of the indole-imidazole heterodimer. They described the V-shaped structure of the ground state complex mainly held by an NH···N hydrogen bond between the NH of indole and the free N of imidazole. They found that the electronic excitation of this compound led to a structural modification corresponding to a tighter binding of the two aromatic groups. For both histidine and tryptophan, ionization leads to a dramatic change of their structures and intrinsic properties.[28-30]

In this study we used cold superfluid helium droplets as an inert matrix to synthesize histidine-tryptophan heterodimers. Embedded in the interior of droplets composed of $\sim 10^5$ helium atoms, the complexes are ionized via a charge transfer from a $He_2^+$ hole migrating through the droplet. To investigate the specific NH···N bond occurring between the two chromophores, imidazole and indole, in an ionized configuration, we studied the reactivity of histidine with the following candidates: tryptophan, indole and methyl-tryptophan.

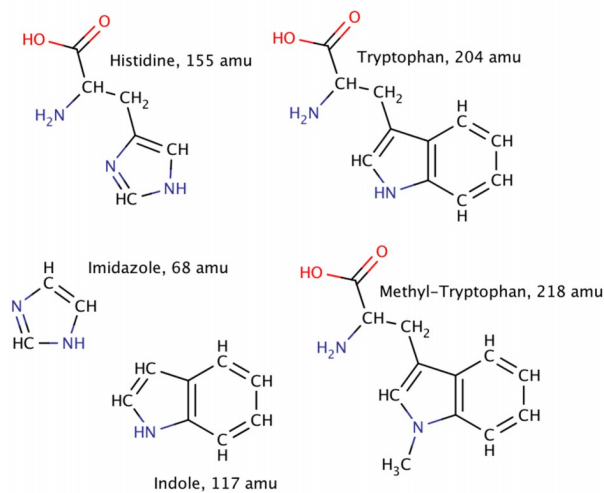

**Fig. 1**. Structures of relevant species.

## II. EXPERIMENTAL SETUP

As illustrated in Fig. 2, experiments were performed using a supersonic beam of $^4$He$_n$ droplets produced by a 5 μm nozzle, cooled to 12 K and with a stagnation pressure of 40 bar. These conditions correspond to an average droplet size of ~10$^4$-10$^5$ atoms.[31] Two consecutive pickup cells are filled with a few milligrams of reactants and heated to a temperature of 150-200 °C. Indole, having a relatively high vapor pressure, was first set in a small capsule with a 1 mm aperture. This capsule was then installed inside one pickup cell and heated as necessary. The beam of droplets was detected by a Balzers QMG-511 crossed-beam quadrupole mass analyzer (QMA) with an electron impact ionization source set at an accelerating voltage of 80 V. A mechanical chopper in conjunction with a lock-in amplifier was used to extract the He beam signal. For all measurements histidine was placed in the first pickup cell in order to ensure a constant pickup rate for this molecule.

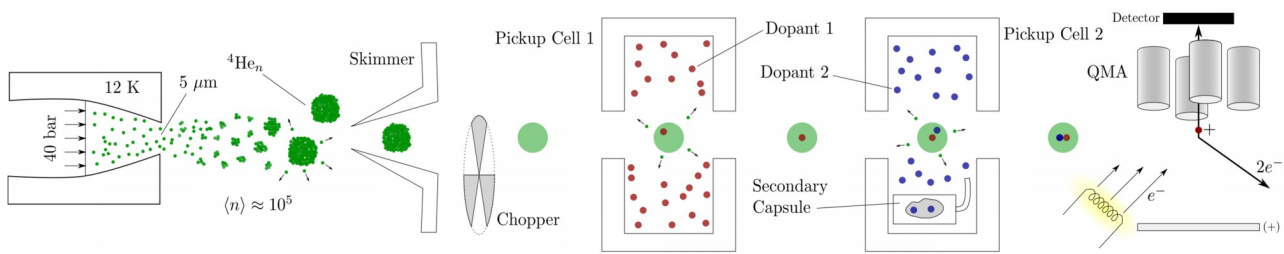

**Fig. 2.** Experimental apparatus used to study doped helium nanodroplets.

## III. RESULTS

### A. Pickup of Histidine

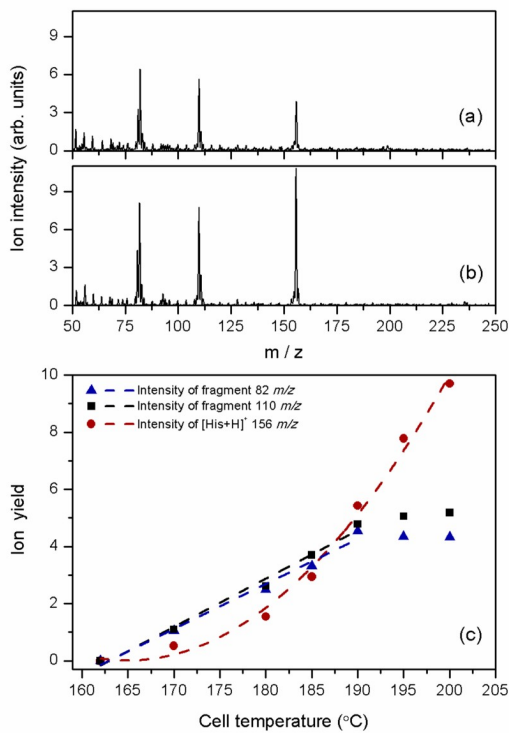

**Fig. 3**. a) Low temperature (170 °C) mass spectrum of histidine doped helium droplets, b) high temperature (190 °C) mass spectrum and c) electron impact ionization yields of the two main fragments (82 and 110 m/z) and [His-H]$^+$ (156 m/z) as a function of pickup cell temperature. The dashed lines represent linear and quadratic fits.

The primary focus of this work is to study the interaction of single histidine molecules with single tryptophan, methyl-tryptophan and indole molecules. Hence, we first determined the temperature of the pickup cell that corresponds to the addition of a single histidine molecule into each helium droplet, on average. As the cell temperature increased, the yield of histidine products exhibited a quick rise between 160 °C and 170 °C. Fig. 3a shows the pickup of histidine with a cell temperature of 170 °C. We mainly observed the two fragments of ionized histidine at 82 and 110 m/z, characteristic of the gas phase.[32] Along with these two ions, we observed only a small intensity of ionized histidine (His$^+$) at 155 m/z (less than 15% of the intensity of fragments), while that of protonated histidine [His+H]$^+$ at 156 m/z is 60%-70% of the fragment peaks' intensity. Fig. 3b shows the pickup of histidine with the cell at 190 °C. We observed that the major product is the [His+H]$^+$ species at this temperature. Fig. 3c shows the intensity of the main peaks observed in the histidine mass spectrum as a function of the cell temperature. The acquisition was stopped as soon as we observed a substantial decrease in the He beam intensity (gauged by the He$_3^+$ signal). The His$^+$ intensity is omitted as it was much lower and its global behavior was similar to the fragment ions.

We clearly see two distinct trends. Firstly, the intensity of the two fragments increased linearly with the temperature and stabilized beyond 190 °C. Secondly, the intensity of [His+H]$^+$ increased quadratically with temperature. We observed that the intersection of those curves coincides with the stabilization of the yield of the fragments. These observations are consistent with previous work[33] and result from an interaction between multiple molecules in each droplet. The [His+H]$^+$ ions are therefore produced from dimers, trimers and larger histidine clusters that fragment upon ionization. Below the cell temperature of 190 ºC, many helium droplets contain single histidine molecules and, consequently, ionization produces more single fragments than [His+H]$^+$ in the mass spectra. Accordingly, we set the histidine cell to a temperature around 170 ºC for the remainder of the experiments.

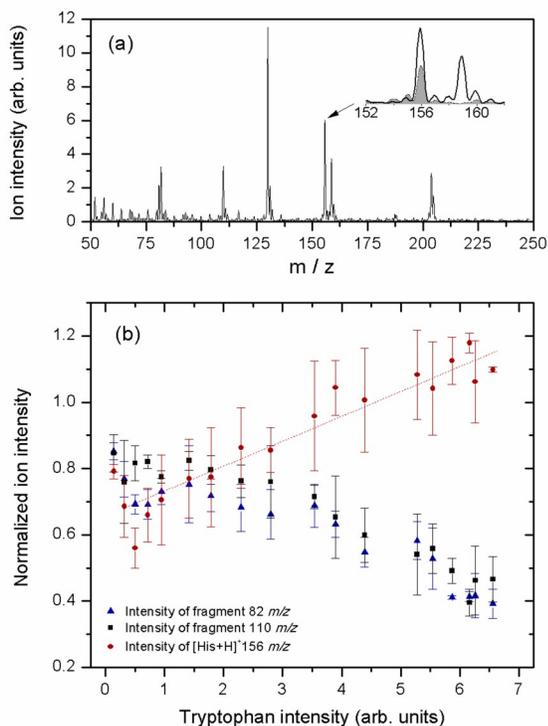

**Fig. 4.** a) Mass spectrum with both histidine and tryptophan. The insert shows the [His+H]$^+$ peak at 156 *m/z* without (gray) and with (solid line) tryptophan. (The peak at 159 *m/z* is a Trp fragment) b) Quantities of ionization products as a function of the intensity of the tryptophan ions (sum of 204, 159 and 130 *m/z* peaks). The plotted data is a moving average (with resulting error bars shown) of a larger set of data, normalized to the intensity of the He$_3$$^+$ peak, and the line is a guide to the eye.

**B. Pickup of Histidine and Tryptophan**

Fig. 4a shows the mass spectra obtained from the pickup of histidine and tryptophan. The spectrum in Fig. 4a was obtained by setting the temperature of the tryptophan pickup cell to 180 C. The mass spectrum of tryptophan is similar to the one presented in the literature.[32, 34] We observed the ionized tryptophan at 204 *m/z* and the characteristic fragments at 130 and 159 *m/z*, produced by electron impact. We observed histidine products similar to those in Fig. 3a, at 82, 110 and 156 *m/z*.

However, upon adding tryptophan the quantity of protonated histidine increased dramatically, while the amount of histidine fragments was likewise reduced. Fig. 4b shows the intensity of the main ions (82, 110 and 156 *m/z*) observed in the histidine mass spectrum as a function of the quantity of tryptophan picked up by the droplets. Their intensities are normalized by the He trimer (He$_3$$^+$) intensity. We adjusted the quantity of tryptophan added to the droplets by varying the tryptophan cell temperature between 140 and 190 C. When the tryptophan cell was removed from the He beam, the [His+H]$^+$ signal reverted to its original intensity, seen in Fig. 3a.

Furthermore, the quantity of histidine fragments decreased by about 50% for the maximum amount of tryptophan. This decrease is likely due to the combination of two effects. With a higher vapor pressure in the tryptophan pickup cell, the number of droplets transmitted through the cell decreases, resulting in a few percent overall decrease of spectrum intensity. Secondly, assuming that the rate of droplet ionization in the QMA is roughly constant (aside from this small decrease due to increased tryptophan vapor pressure), adding tryptophan (or any other dopant) to each droplet decreases the probability of ionizing the embedded histidine. However, by ionizing tryptophan in the droplet, a proton transfer from Trp$^+$ to histidine becomes more probable, leading to an increase in the intensity of [His+H]$^+$. Moreover, for fixed quantities of histidine, this increase would be proportional to the amount of tryptophan in the droplets, which is what we observed.

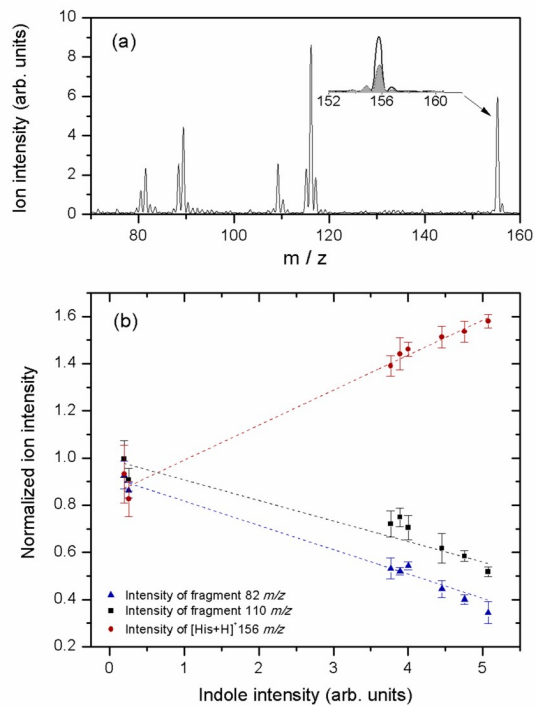

**Fig. 5:** a) Mass spectrum with both histidine and indole. The insert shows the [His+H]$^+$ peak without (gray) and with (solid line) indole. b) Relative quantities of ionization products as a function of the intensity of the indole ion (116 *m/z*). The plotted data is a moving average (with resulting error bars shown) of a larger set of data, normalized to the intensity of the He$_3$$^+$ peak. The dashed lines are linear fits.

## C. Pickup of Histidine and Indole

Fig. 5a shows the mass spectra corresponding to the pickup of histidine and indole. Due to the high vapor pressure of indole, even at room temperature (25 ºC), it was difficult to limit the quantity of indole inserted into the droplets. For this reason, large quantities of indole were picked up with its cell at room temperature, and there are no data points for intermediate quantities of indole. In future work, it may be possible to overcome this difficulty by cooling the cell. The mass spectrum of indole was similar to that obtained in the gas phase.[32] We observed a large peak at 116 $m/z$ corresponding to ionized indole as well as the histidine ions described above. Compared to the histidine spectrum (Fig. 3a), the difference between the amounts of protonated histidine and the fragments is, as with tryptophan, reversed. Fig. 5b shows the intensity of the main peaks observed in the histidine mass spectrum as a function of the amount of indole ion. Similar to the histidine-tryptophan system, we observed that the intensity of [His+H]$^+$ signal increases linearly with the indole signal intensity, for the range of acquired data. The quantity of histidine fragments is reduced as well.

This suggests that a proton is transferred from the indole group to histidine upon ionization of the heterodimer and not from the backbone of the amino acid.

## D. Pickup of Histidine and Methyl-Tryptophan

Fig. 6a shows the mass spectrum corresponding to the pickup of histidine and methyl-tryptophan. The spectrum was obtained by heating the methyl-tryptophan cell to a temperature of 180 C. The only difference between this scenario and the one with tryptophan above is the methylation of the amino group of the indole side chain, which prohibits the N-H interaction between side chains. We again observed the characteristic fragments of ionized histidine (cf. Fig 3a). The mass spectrum observed for methyl-tryptophan alone was identical to that of tryptophan but shifted by a mass of 14, due to the methyl group.

Fig. 6b shows the intensity of the main peaks observed in the histidine mass spectrum as a function of methyl-tryptophan picked up by the droplets. As we observed with tryptophan and indole, the signal intensity of the histidine fragments was decreased, mostly due a reduced probability of ionizing a histidine molecule in each droplet by adding another dopant. But in this case the quantity of the [His+H]$^+$ ion did not change significantly. In contrast to the previous scenarios with tryptophan and indole, we observed that the quantity of [His+H]$^+$ remained stable as the quantity of methyl-tryptophan increased. The lack of protonation in this case is yet further evidence that the proton transferred to histidine from tryptophan originates from the NH group of the indole side group.

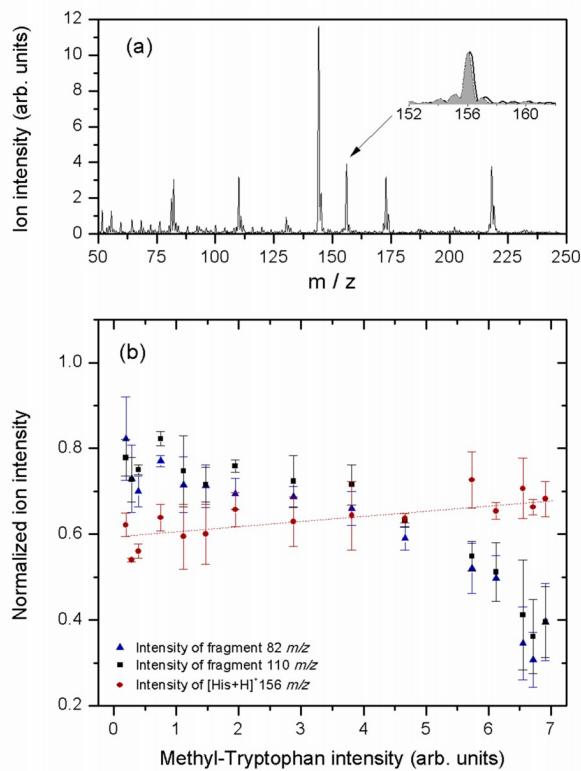

**Fig. 6:** a) Mass spectrum with both histidine and methyl-tryptophan. The insert shows the [His+H]$^+$ peak without (gray) and with (solid line) methyl-tryptophan. b) Relative quantities of ionization products as a function of the intensity of the methyl-tryptophan ions (sum of 218, 173 and 144 $m/z$ peaks). The plotted data is a moving average (with error bars shown) of a larger set of data, normalized to the intensity of the He$_3^+$ peak, and the line is a guide to the eye.

## Discussion

We have observed that the formation of tryptophan-histidine complex can lead to the formation of a protonated histidine [His+H]$^+$ upon ionization. Experimental evidence has allowed us to assign this protonation to be specific to the presence of the indole group. In addition, to rule out protonation of the amino acid backbone, we combined tryptophan and diglycine in helium droplets and saw no protonation of diglycine due to tryptophan.

While there is no direct evidence that the proton transfer occurs after or before the ionization of the histidine-tryptophan complex embedded in helium droplets, it is unlikely that such a reaction would occur at 0.37 K between neutral amino acids, both with high proton affinities.[35]

Assuming that the proton transfer occurs after ionization,

a major challenge is to understand which one within the complex, histidine or tryptophan, is ionized first. Upon ionization of the droplet, a hole ($He_2^+$) forms and migrates in a random walk through the droplet before localizing on the embedded species.[31, 36, 37] The overall trajectory of the migrating hole is strongly influenced by the long-range electrostatic fields in the droplet. In the case of a dopant with a strong dipole moment, the hole is attracted to its negative pole and repelled by its positive pole.[38, 39] Both histidine and tryptophan have relatively large dipole moments. A measurement of the dipole moment of the complex may be an appropriate way to address this question.

Using semi-empirical and density functional theory (DFT) calculations,[40-42] we found that the V-shaped minimum energy imidazole-indole complex identified and observed by Kumar et al.[27] has its dipole moment directed from the indole to the imidazole group (see Fig. 7). This preliminary calculation suggests that the tryptophan residue attracts the hole and accepts the positive charge. A proton is then donated from a π-cation radical tryptophan to the imidazole group, guided by the hydrogen bond.

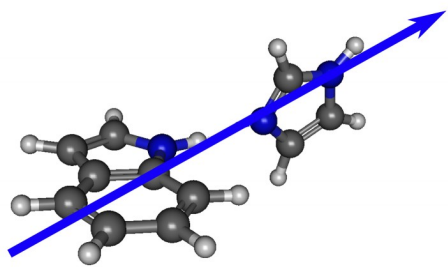

**Fig. 7.** Dipole moment and structure of the hydrogen bonded indole-imidazole complex based on Ref. 27. The structure shown in the insert was generated manually, optimized with the semi-empirical computational chemistry package MOPAC2012[40] with PM6-DH+ methodology, and further optimized with DFT using the B3LYP functional and the 6–311G* basis[41]. The Gabedit software package [42] was used for data analysis and visualization. The dipole moment of the minimal energy structure, calculated to be 7.8 D, is oriented in the indole-imidazole direction. For validation, the vibrational frequencies of this indole-imidazole complex, computed with B3LYP/6-311G*, were found to match the experimentally measured values,[27] lending support to the calculation of its bond structure and dipole moment.

## Conclusions

Aromatic-aromatic interactions are important in the ultimate configurations of polypeptides. We observed the interaction between the representative amino acids, histidine and tryptophan, in nanoscale superfluid $^4$He droplets. Upon ionization of the histidine-tryptophan heterodimer, we found clear mass spectrometric indications of a proton transfer from tryptophan to histidine. Furthermore, by also combining histidine with indole and methyl-tryptophan inside the droplets, we showed that the proton transfer occurs across the NH···N hydrogen bond between the indole and imidazole side chains.

Research towards the measurement of dipole moments of such complexes embedded in helium droplets is currently under development in our group. It is also interesting to consider analogous investigations involving other aromatic amino acid residues as well as peptides.

## Acknowledgments
We would like to thank Kayla Mendel for her assistance during experiments and Daniel Ben-Zion for his technical help. This research was supported by the U.S. National Science Foundation under grant CHE-1213410.

* Present address: The University of Manchester, Manchester, M1 7DN, United Kingdom